\documentclass[10pt,letterpaper]{article}
\usepackage[top=0.85in,left=1.25in,footskip=0.75in]{geometry}

\usepackage{changepage}

\usepackage[utf8]{inputenc}

\usepackage{textcomp,marvosym}

\usepackage{fixltx2e}

\usepackage{amsmath,amssymb}

\usepackage{cite}

\usepackage{nameref,hyperref}

\usepackage[right]{lineno}

\usepackage{microtype}
\DisableLigatures[f]{encoding = *, family = * }

\usepackage{rotating}

\usepackage{setspace} 
\raggedright
\usepackage[aboveskip=1pt,labelfont=bf,labelsep=period,justification=raggedright,singlelinecheck=off]{caption}

\usepackage{verbatim}
\usepackage[normalem]{ulem}
\usepackage[usenames,dvipsnames,svgnames,table]{xcolor}

\makeatletter
\renewcommand{\@biblabel}[1]{\quad#1.}
\makeatother

\date{}

\usepackage{lastpage,fancyhdr,graphicx}
\usepackage{epstopdf}
\newcommand{\edit}[1]{{#1}}
\renewcommand{\sout}[1]{}

\begin{document}
\vspace*{0.35in}

\begin{flushleft}
{\Large
\textbf\newline{Has wild poliovirus been eliminated from Nigeria?}
}
\\
{\large July 10, 2015: updated forecast for the ongoing circulation of type 2 vaccine-derived polio}
\newline
\\
Michael Famulare\textsuperscript{1,*}
\\
\bigskip
\bf{1} Institute for Disease Modeling, Bellevue WA, USA
\\
\bigskip

%
%





* mfamulare@intven.com

\end{flushleft}
\section*{Abstract}

Wild poliovirus type 3 (WPV3) has not been seen anywhere since the last case of WPV3-associated paralysis in Nigeria in November 2012. At the time of writing, the most recent case of wild poliovirus type 1 (WPV1) in Nigeria occurred in July 2014, and WPV1 has not been seen in Africa since a case in Somalia in August 2014. No cases associated with circulating vaccine-derived type 2 poliovirus (cVDPV2) have been detected in Nigeria since November 2014. Has WPV1 been eliminated from Africa? Has WPV3 been eradicated globally? Has Nigeria interrupted cVDPV2 transmission? These questions are difficult because polio surveillance is based on paralysis and paralysis only occurs in a small fraction of infections.

This report provides estimates for the probabilities of poliovirus elimination in Nigeria given available data as of March 31, 2015. It is based on a model of disease transmission that is built from historical polio incidence rates and is designed to represent the uncertainties in transmission dynamics and poliovirus detection that are fundamental to interpreting long time periods without cases.

The model estimates that, as of March 31, 2015, the probability of WPV1 elimination in Nigeria is $84\%$, and that if WPV1 has not been eliminated, a new case will be detected with $99\%$ probability by the end of 2015. The probability of WPV3 elimination (and thus global eradication) is $>99\%$. However, it is unlikely that the ongoing transmission of cVDPV2 has been interrupted; the probability of cVDPV2 elimination rises to $83\%$ if no new cases are detected by April 2016.

{\bf Added July 10, 2015}: On June 26, 2015, a paralytic polio case caused by ongoing circulation of established cVDPV2 lineage was confirmed by the Global Polio Laboratory Network. The date of paralysis onset was May 16, 2015, 181 days after the previous case. The case provides new information about the prevalence of cVDPV2 in Nigeria that can be used to both assess the accuracy of the predictions for elimination and case detection in the first version of the paper and provide the basis for an updated forecast for the time to cVDPV2 elimination if no cases are detected.  

The model predictions for cVDPV2 pepared on March 31, 2015 proved to be accurate: elimination did not occur and the observed case occurred within the $95\%$ prediction interval. The updated forecasts predict that this one case has little impact on the estimates for the probability of elimination by April 2016. This counter-intuitive result was found because the 6-month silent interval following the previous 13-day interval indicates that cVDPV2 prevalence has fallen roughly 16-fold in Nigeria since November 2014 and that the recent mean effective reproductive number for ongoing cVDPV2 transmission in Nigeria is below 1. Under the most likely scenario for the transmission dynamics in the near future, the median time to the date of onset of the next case if elimination does not occur is July 15, 2015. 

I intend to provide further updates to this manuscript if new cases occur. It is important to note that if another case occurs before the end of July 2015, it will likely indicate increasing prevalence and so forecasts for elimination by April 2016 would then be more pessimistic.

\section*{Introduction}
Since the World Health Assembly announced the intention to eradicate polio worldwide in 1988, all countries except Pakistan, Afghanistan, and Nigeria have \sout{certified elimination}\edit{interrupted indigenous transmission} of all three wild poliovirus (WPV) serotypes at least once \cite{Kew2014}. Nigeria may be the next country to achieve wild-type polio elimination. The most recent case of wild-type 1 (WPV1) in Nigeria occurred in July 2014 \cite{GPEI2015}. The most recent case of wild-type 3 (WPV3) seen globally occurred in Nigeria in November 2012 \cite{Kew2014}. Is Nigeria wild-polio-free? Is WPV3 eradicated globally? Furthermore, the last case of WPV1 in Africa occurred in Somalia at the tail end of the Horn of Africa Outbreak in August 2014. Is Africa wild-polio-free?

The primary mechanism for detecting poliovirus in a population is surveillance for paralytic poliomyelitis. The incidence of poliomyelitis is tracked by the acute flaccid paralysis (AFP) surveillance and global polio laboratory network \cite{WHOAFPSurveillance,Hull1997}. The certification period to declare that a polio serotype has been eliminated from a previously endemic region with certification-standard surveillance is three or more years without a polio case\edit{ associated with indigenous WPV circulation}\cite{Smith2004a}. This criterion is based on experience with regional elimination \cite{Debanne1998a} and is necessarily long because paralysis due to polio is uncommon. Estimated case-to-infection ratios\edit{ in fully-susceptible individuals} are 1:200 for wild type 1, 1:1900 for wild type 2, and 1:1150 for wild type 3 \cite{Nathanson2010}.  Previous modeling work has supported that the three year rule of thumb is reasonable, but that more precise estimates of the elimination time depend on the case-to-infection ratio for each serotype and the local conditions leading up to elimination \cite{Eichner1996,Kalkowska2012,Kalkowska2015}. 

Polio elimination in the most challenging settings has been driven by supplemental immunization activities that require extensive resources and prolonged community engagement to be successful \cite{Sutter2006,Bahl2014,GPEI2015}. Operational capacity and community engagement can be difficult to maintain for years in the presence of political instability, humanitarian crises, and substantial health care needs unrelated to polio \cite{Force2013,GPEI2015}. While it is reasonable to maintain the three year certification period as the gold standard, the operational, humanitarian, and financial requirements support the need for specific, data-driven estimates of the likely period of silent polio persistence.

The need for specific silent duration estimates is especially important in the context of the plan to stop all use of trivalent oral polio vaccine (tOPV) in April 2016 and replace it with bivalent OPV (bOPV) containing only serotypes 1 and 3 to prevent the seeding of new type 2 circulating vaccine-derived polio (cVDPV2) outbreaks \cite{WHOSAGEPolioWorkingGroup2014,Kew2005a}. cVDPV outbreaks can occur when vaccine strains transmit for sufficient time to allow for genetic reversion of the markers of attenuation \cite{Kew2005a}. OPV2 is the most common source of cVDPV globally \cite{DuintjerTebbens2013}, and Nigeria has had ongoing transmission of cVDPV2 since 2005 \cite{Burns2013a,GPEI2015}. In 2014, cVDPV2 was the dominant cause of poliomyelitis in Nigeria \cite{GPEI2015}, and this is likely because Nigeria has been primarily using bOPV to drive the successes in WPV control \cite{Etsano2014,Mangal2014}. However, due to expanded use of tOPV and combined tOPV/inactivated polio vaccine campaigns from the second half of 2014 \cite{IMB2014,Etsano2014}, cVDPV2 cases in Nigeria abruptly stopped appearing at the end of November 2014 after producing 30 cases earlier that year \cite{GPEI2015}. Are the established cVDPV2 lineages gone? Will they be gone before the planned April 2016 tOPV cessation deadline? When is it reasonable for Nigeria to switch from a strategy focused on WPV elimination to one focused on cVDPV2? 

To estimate the probabilities of elimination from Nigeria given the data available at the time of preparation, March 31, 2015, I built a model appropriate to answer the question: how long do we have to wait before a chain of transmission either terminates or produces another case? The model is based on four premises: (1) that elimination only depends on the last few hundred to few thousand infections in a much larger population, (2) that we only need to see one more case to know elimination did not occur, (3) that we do not care where infected people are located because AFP surveillance will find any cases\edit{ in proportion to the surveillance sensitivity}, and (4) that the few remaining pockets of infection have similar transmission dynamics to the many sources of polio cases in the past.   The model incorporates both the parametric and stochastic uncertainties that limit our knowledge of polio transmission. It provides estimates of the probability of elimination given no new cases and the time to the next case if elimination does not occur, and the results are appropriate for supporting the rational analyses \cite{Hansson2005} of planned polio vaccination policies. 

{\bf Added July 10, 2015}: On June 26, 2015, a paralytic polio case caused by ongoing circulation of established cVDPV2 lineage was confirmed by the Global Polio Laboratory Network. The date of paralysis onset was May 16, 2015, 181 days after the previous case. With this new information, I assess the accuracy of the predictions for elimination and case detection in the first version of the paper and provide an updated forecast for the time to cVDPV2 elimination if no cases are detected.  New sections are marked by the ``Added July 10, 2015'' tag.

\section*{Results}
\begin{figure}[h]
\includegraphics{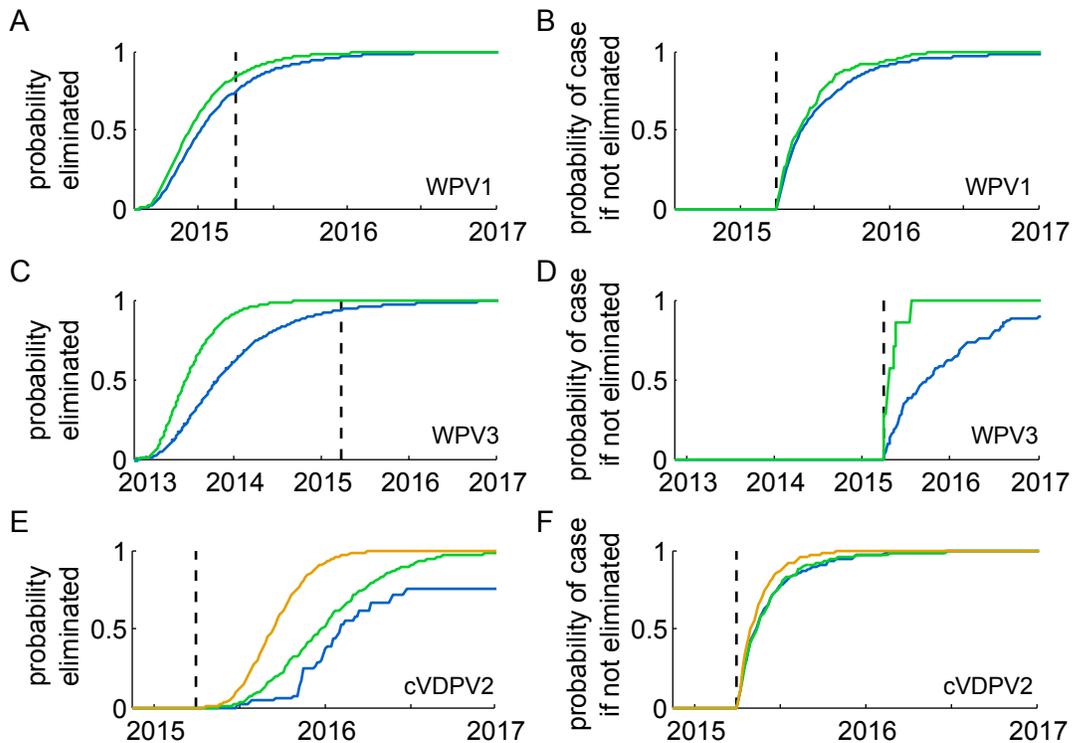}
\caption{{\bf Model results for elimation and time to next case.} Probability of elimination assuming no new cases are observed: WPV1 (A), WPV3 (C), cVDPV2 (E). Probability of observing a new case if elimination doesn’t occur, WPV1 (B), WPV3 (D), cVDPV2 (F). Fig. prepared March 31, 2015 (black dashed line). In each panel, the horizontal axis origin is the time of the most recent case, the\edit{ solid} blue curve shows the\sout{ worst-case} scenario with mean $R_{\text{eff}}=1$\edit{ and perfect surveillance}, \sout{and }the green curve shows a less conservative scenario with mean $R_{\text{eff}}<1$\edit{ and perfect surveillance, and the dashed blue curve shows a conservative scenario with mean $R_{\text{eff}}=1$ and $50\%$ surveillance sensitivity}. For cVDPV2 (C,F), the orange curve depicts an optimistic scenario in which the mean $R_{\text{eff}}$ is held at the lowest value ever observed and the standard deviation is reduced to one-fourth of its observed value. }
\label{resultsFig}
\end{figure}

The left column of Fig. \ref{resultsFig} depicts elimination: it shows the probability by date that each serotype been eliminated if no new cases are detected. The right column depicts case detection: it shows the probability of the time to the next case if elimination does not occur\sout{ under the assumption that surveillance is of sufficiently high quality that no polio cases are missed}. The dashed line indicates March 31, 2015. The\edit{ solid} blue curves provide estimates based on models with mean effective reproductive number equal to 1 (mean growth rate of the number of active infections equal to 0).\edit{ For a fixed surveillance sensitivity,} $R_{\text{eff}}=1$ is the worst-case scenario because it maximizes the possible duration of silent transmission without elimination. Deviations from this assumption either make elimination easier $(R_{eff}<1)$ or make the silent period between cases shorter $(R_{eff}>1)$. For WPV1 and WPV3, the green curves provide less conservative estimates based on the assumption that vaccination quality maintains mean $R_{\text{eff}}<1$ (growth rate $<0$) at the level estimated at the time of the last case. For cVDPV2, the green curve assumes that the mean growth rate is held at one standard deviation below zero. Comments below summarize results from the green curves unless otherwise stated. \edit{For WPV1 and WPV3, the dashed blue curves provide conservative estimates under the assumptions that $R_{\text{eff}}=1$ and surveillance is only likely to detect $50\%$ of polio cases. The poor surveillance curves are not shown for cVDPV2 because elimination is very unlikely, occurring in less than 1 in 100~000 simulations. }

\paragraph*{WPV1.}
The most recent WPV1 case in Nigeria occurred on July 24, 2014. As of March 31, 2015, there is an $84\%$ chance that WPV1 has been eliminated from Nigeria. The chance of elimination will be \sout{$99\%$}\edit{$93\%$} by the end of 2015 if no new cases are seen even under the worst-case scenario for\edit{ surveillance and} the force of infection (Fig. \ref{resultsFig}A, dashed blue curve). If WPV1 does not eliminate before making another case, there is a $66\%$ chance we will see a new case by the end of June 2015 and a $95\%$ chance by the end of 2015\edit{ (Fig. \ref{resultsFig}B)}. 

The most recent case outside of Nigeria occurred in Somalia on August 11, 2014 at the tail end of the Horn of Africa outbreak that begin in 2013, 69 days after the case preceding it \cite{GPEI2015a}. The corresponding predictions from the model for Somalia are thus essentially identical to those for Nigeria. The combined probability that both countries have eliminated (and thus all of Africa) is the square of the Nigeria values, and so the model estimates a $70\%$ chance that Africa is wild-polio-free as of March 31, 2015. 

\paragraph*{WPV3.}
The most recent WPV3 case globally occurred on November 11, 2012 in Nigeria. As of March 31, 2015, there is a $>99\%$ chance that WPV3 has been eliminated from Nigeria\sout{ and thus eradicated globally}.\edit{ The most recent case detected anywhere outside of Nigeria was in Pakistan seven months earlier (April 2012). The detection of the other serotypes since 2012 in both (and neighboring) countries indicates that surveillance is able to detect WPV3 if it were present, and so elimination from Nigeria likely represents global eradication.}  In the unlikely event WPV3 has not been \sout{eradicated}\edit{eliminated}, there is a $>99\%$ chance we will see a new case by the end of 2015. 

\paragraph*{cVDPV2.}
The most recent cVDPV2 case in Nigeria occurred on November 16, 2014 from a genetic lineage that has been in circulation since 2005. As of March 31, 2015, it is unlikely that the known cVDPV2 lineages \cite{Burns2013a,GPEI2015} have been eliminated from Nigeria. 

If the known cVDPV2 lineages are still in circulation, there is an $84\%$ chance the next case will appear by the end of June 2015, and a $97\%$ chance a case will appear by the end of 2015. This model prediction for silent persistence is consistent with the longest observed interval of approximately 7 months seen from November 2012 to June 2013 \cite{Etsano2014,Burns2013a,Diop2014}. 

The prediction for the probability of cVDPV2 elimination is sensitive to assumptions about future type 2 immunity. The rapid disappearance of cases in late 2014 (from one case every 10 days on average in 2014 to no cases in over 4 months) implies that type 2 immunity is higher in the relevant population than it has been at any time previously, and so models based on historical transmission rates are likely too conservative. As population immunity affects transmission dynamics through herd effects that reduce the force of infection \cite{Fine2011}, consider three scenarios for the force of infection in 2015 and beyond:
\begin{itemize}
	\item Historically typical tOPV coverage (blue):\edit{ For the historically typical coverage scenario, the model assumes that the future force of infection is described by the historical estimates---the mean growth rate is zero ($R_{\text{eff}}=1$) and the standard deviation is described in the Methods. Under this scenario}\sout{ If the mean growth rate is zero ($R_{\text{eff}}=1$)}, elimination is unlikely with a $1\%$ chance prior to June 2015. The estimated probability of elimination without seeing a case rises to $66\%$ by April 2016, but does not reach $95\%$ until June 2017. 
	\item Good tOPV coverage (green):\edit{ For a good coverage scenario, the model assumes that tOPV coverage improvements will maintain a reduction in the mean force of infection to}\sout{ If growth rates persist at} one standard deviation below the historical norm \sout{but}\edit{and so} not at unprecedented levels, $\lambda-\mu = -3.44 \,\text{yr}^{-1}$ (such that mean $R_{eff}\approx 0.85$).\edit{ Under this scenario,} there is only a $3\%$ chance of elimination by the end of June 2015 and an $83\%$ chance that cVDPV2 will be eliminated by April 2016 if no new cases are seen.  The probability of elimination rises to $98\%$ by the end of 2017. 
	\item Excellent tOPV coverage (orange):\edit{ For the excellent coverage scenario, the model assumes that tOPV coverage improvements will maintain the mean growth rate at the lowest level transiently observed in the last ten years, $\lambda-\mu = -9.21 \,\text{yr}^{-1}$ (such that mean $R_{eff}\approx 0.60$), and will also reduce the variability of the growth rate to one-fourth of the historical standard deviation.}\sout{ If the recent improvements in tOPV vaccination quality and higher tOPV campaign frequency maintain the mean growth rate at the lowest level observed in any interval for any serotype in the last ten years $\lambda-\mu = -9.21 \,\text{yr}^{-1}$ (such that mean $R_{eff}\approx 0.60$) and reduce the variability of the growth rate to one-fourth of the historical standard deviation,}\edit{ Under this scenario}, there is a $13\%$ chance of elimination by the end of June 2015 and a $99\%$ chance cVDPV2 will be eliminated by April 2016 if no new cases are seen.  
\end{itemize} 

This analysis is focused solely on estimating the elimination probability for established cVDPV2 lineages and does not account for the emergence of new cVDPV2 lineages. In the event of a new emergence, these estimates would remain valid for the current genetic lineages, but additional modeling would be required to produce estimates for any new lineages and thus cVDPV2 overall.

\begin{figure}[h!]
\includegraphics{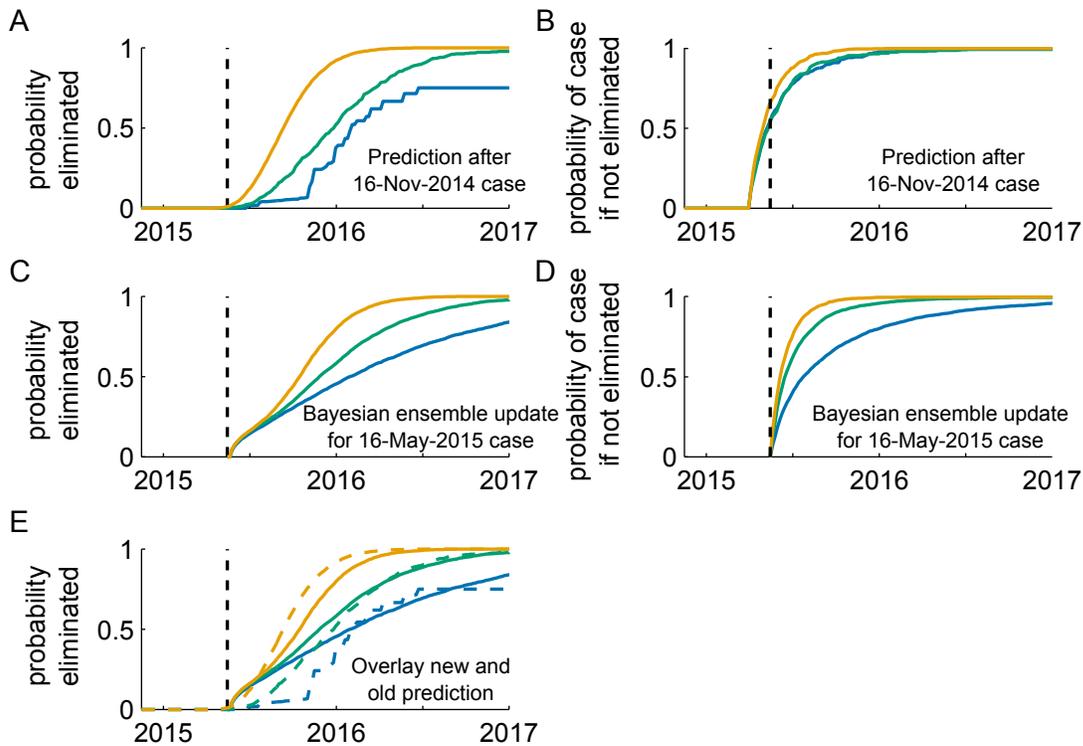}
\caption{{\bf Comparison of observed May 16, 2015 cVDPV2 case with the original forecast, and updated forecasts.} The dashed black line marks the date of onset of the most recent cVDPV2 case on May 16, 2015. (A) Original forecast from March 31, 2015 for the probability of cVDPV2 elimination assuming no new cases are observed and (B) the probability of observing a new case if elimination doesn’t occur (reproduced from Fig. \ref{resultsFig}). (C) Updated forecast prepared on June 30, 2015 for elimination if no new cases are observed and (D) probability of the time to the next case if elimination does not occur. (E) Overlay of new forecast (solid lines) and original forecast (dashed lines) for easier comparison. As in Fig. \ref{resultsFig}, the\edit{ solid} blue curve shows the conservative scenario with mean $R_{\text{eff}}=1$\edit{ and perfect surveillance}, \sout{and }the green curve shows the scenario most consistent with the recent 6-month silent period with mean $R_{\text{eff}}<1$\edit{ and perfect surveillance,} and the orange curve depicts an optimistic scenario in which the mean $R_{\text{eff}}$ is held at the lowest value ever observed and the standard deviation is reduced to one-fourth of its historical value. Starting from May 16, $31\%$ of all simulations in the blue scenario eliminate without producing another case, $61\%$ eliminate in the green scenario without a case, and $73\%$ eliminate in the orange without a case. }
\label{JulyResultsFig}
\end{figure}

{\bf Added July 10, 2015:} The new cVDPV2 case on May 16, 2015 provides information to validate the previous model and to update the cVDPV2 forecast. The original forecast was accurate in that it both correctly predicted that elimination was very unlikely prior to May 2015 (Fig. \ref{JulyResultsFig}A), and the observed case occurred at the $54^{\textrm{th}}$-percentile of the prediction interval for the time to the next case given the data available as of March 31, 2015 (Fig. \ref{JulyResultsFig}B).

For each of the three scenarios considered for the original forecast, I estimated the probablility that each scenario produced the observed outcome. For the historically typical tOPV coverage scenario (blue), the observed outcome of a case after 181 days occured in 8 of 100~000 simulations; for the good tOPV coverage scenario (green), 23 of 100~000 simulations; and for the optimistic scenario (orange), 4 of 100~000 simulations. The absolute probabilites are low because any specific point outcome is unlikely, but the differences in the probabilities provide evidence to weight how consistent each scenario is with the observed outcome. That the green scenario is most likely to produce the observed outcome supports the assumption of the original paper that a scenario in which the mean $R_{eff}$ is below 1 but not at unprecedentedly low levels is reasonable to describe the present situation. 

As described in the Methods, I used the set of previous simulations consistent with the observed outcome to estimate the prevelance of cVDPV2 infection in Nigeria on May 16. The median estimated infection prevalence on May 16, 2015 is 103 infections, which is roughly 16-fold lower than the median estimated prevelance in November 2014. Thus, the new case, while unfortunate, provides evidence of significant reductions in cVDPV2 prevalence. 

From the estimated initial condition distribution, the model produces updated forecasts from May 16, 2015 for the probability of elimination given no new cases (Fig. \ref{JulyResultsFig}C) and the time to the next case if elimination does not occur (Fig. \ref{JulyResultsFig}D). Under the most likely of the three simulated scenarios for $R_{eff}$ in which the mean effective reproductive number is one standard deviation below the historical norm (green), the probability of elimination if no new cases occurs is $64\%$ by the end of 2016, $85\%$ by April 2016, and $99\%$ by the end of 2016. These results are essentially unchanged from the estimate prepared on March 31 (Fig. \ref{JulyResultsFig}E). The model actually increases its estimate of the probability of elimination in the next few months because the new case reduces the uncertainty in the space of possible trajectories that allow for persistance through May without cases or elimination. If elimination does not occur, the median predicted date of onset of the next case is July 15, and there is a $92\%$ chance of another case by the end of the year (Fig. \ref{JulyResultsFig}D).

\section*{Discussion}

As of March 31, 2014, the model predicts that WPV3 has been eradicated globally and there is an $70\%$ chance that Africa is completely wild-polio free for the first time in history. Furthermore, in the now unlikely situation that WPV1 has not yet been eliminated from Nigeria and that it persists with the worst-case mean reproductive number of exactly 1\edit{ and only $50\%$ surveillance sensitivity}, we can expect to see it or for it to eliminate with \sout{$95\%$}\edit{$93\%$} certainty by the end of 2015. Thus, this author believes it will be reasonable to conclude that Africa is wild-polio free if no new WPV cases are seen by the end of 2015. This estimate for time to elimination of at most 14 months is significantly shorter than estimated in previous work \cite{Eichner1996,Kalkowska2012,Kalkowska2015} because the model takes into account the low infection prevalence at the time of the most recent case that is implied by the long interval between the most recently observed cases. 

For cVDPV2, despite the dramatic reduction in incidence implied by the lack of new cases since November 2014, we must be cautious about concluding too soon that the established lineages are gone. The low case to infection ratio of type 2 makes it the most difficult strain to detect at low prevalence, and the high rate of incidence in 2014 implies that the reservoir of existing silent infections in November 2014 was approximately 22 times larger for cVDPV2 than for WPV1 at the time of its last case. The larger reservoir delays elimination even with sufficient herd immunity. Furthermore, continued tOPV use may lead to new cVDPV emergences in populations with insufficient herd immunity. The disappearance of cases indicates that substantial improvement in type 2 immunity has significantly reduced the national incidence rate, but the historical stability of the effective reproductive number near 1 suggests that the dynamics in the populations that continue to transmit are more stable than the national trend overall. Estimates here apply only to the known endemic lineages and new emergences may change expectations for the time to complete cVDPV2 elimination even if no new cases are seen from the established lineages.

The above points suggest to this author that it would be reasonable for the polio program in Nigeria to place more emphasis on cVDPV2 prevention by the end of 2015. However, until the world is wild-polio-free, the risk of the re-importation of WPV1 must be balanced with the risk of ongoing cVDPV2 transmission and new cVDPV emergence. Furthermore, certification of polio-free status is a stringent goal \sout{wiith}\edit{with} additional requirements for surveillance and specimen containment \cite{Smith2004a}, and so these estimates need to be considered in that context. The model was designed to help policymakers manage that balance. Involved parties may use the probabilities over time to weight different scenarios and inform their planning \cite{Hansson2005}, and the model can be easily extended to consider alternative scenarios about the future dynamics of the force of infection. 

If a new case is detected, then these model predictions are no longer valid. However, since the silent periods as of March 31, 2015 are much longer than typical, a new case would likely represent one of two scenarios. It would either be the last orphan case very near elimination, or it will be the first case of a new outbreak for which more cases will shortly follow. In the event another case occurs but is not followed by an outbreak, this analysis can be repeated to provide updated estimates. In the event of an outbreak, surveillance sensitivity is not the primary policy concern.

The model assumes that the balance of birth and vaccination maintains stationary dynamics for the force of infection with constant mean infection growth rate and random variability. With additional data about birth demographics and vaccination activities, this model can be extended to include non-random structure in the statistical model of the force of infection. One could inform such models by correlating the apparently random changes in the force of infection with the supplemental vaccination calendar \cite{Sutter2006,Bahl2014,IMB2014,GPEI2015} to estimate how future campaigns may affect the force of infection and subsequent probability of elimination.

The model is based solely on AFP surveillance and ignores any role for alternative active surveillance strategies such as environmental surveillance \cite{Hovi2012}. Certification-quality AFP surveillance provides a representative sample from all populations regardless of location. It is not obvious how to include environmental surveillance (ES) in the model because ES is only sensitive at fixed locations, but the model makes no assumptions about spatial structure.\edit{ Furthermore, at a given ES site, there is not yet any practical way to quantify the meaning of a negative sample. And while positive environmental samples provide unambiguous evidence of local transmission at the time of the sample, we do not yet know how to quantify the effect of a positive sample on quantitative estimates of prevalence and thus future elimination probabilty.} Because ES can dramatically improve the sensitivity in well-characterized populations \cite{Hovi2012,Shulman2014}, the role of ES in certifying elimination warrants further empirical and theoretical study.

{\bf Added July 10, 2015:} \edit{After this paper was submitted, the GPEI reported an environmental sample from March (Kaduna State, Nigeria) that was positive for cVDPV2 from an established lineage \cite{GPEI2015c}, and a new case from the same lineage with a date of onset of May16, 2015 was reported on June 30 in the Federal Capital Territory, confirming the predictions of this paper that cVDPV2 elimination was unlikely at that time and that a new case was likely within a few months of March 31, 2015. } 

The model estimates that this one case does not significantly change the forecast for cVDPV2 elimination by April 2016. This perhaps counter-intuitive result occurs because the new case is consistent with a substantial drop in cVDVP2 prevalence over the last 6 months.  The median predicted time to the next case is 2 months under the most likely scenario for the future effective reproductive number. If another case with a date of onset prior to the end of July 2015 is observed, then it will be likely that cVDPV2 prevalence is rising, and this would lead to significantly more pessimisitc forecasts for cVDPV2 elimination by April 2016. I will continue to provide updated forecasts if new cases occur.

\section*{Methods}
The persistent intervals prior to elimination and the silent periods between cases given no elimination are determined by the force of infection, the case-to-infection ratio for each serotype, the mean infectious duration, and the number of extant infections at the time of the most recent case. In contrast to complex models with many elements that explicitly model assumptions about the influences of demographics, contact patterns, heterogeneous immune states, and vaccination history on transmission \cite{Eichner1996,Kalkowska2012,Kalkowska2015}, this model reduces all the complexity of transmission into a statistical model for the time-varying force of infection. This design facilitates efficient marginalization over the substantial uncertainty in transmission dynamics in a manner consistent with the limited available data. The range of the simulated outcomes from the model honestly represents the contributions of both parametric and stochastic uncertainties to the estimates of the probabilities of elimination. 

\paragraph*{Dynamical model.}
 The deterministic approximation to the model is given by a single equation:
\begin{equation*}
	\dot{N}_i=\left(\lambda(t)-\mu\right)N_i
\end{equation*}
where $N_i$ is the number of infected people at time t, $\lambda(t)$ is the time-varying force of infection, and $\mu$ is the inverse of the mean infectious duration. The growth rate is $\left(\lambda(t)-\mu\right)$ and the effective reproductive number is $R_{\text{eff}} =\frac{\lambda}{\mu}$. In the stochastic version of the model, new infections occur at the inhomogeneous Poisson rate $\lambda(t)N_i(t)$ and existing infections are cleared at Poisson rate $\mu$. 

This simple linear model is appropriate to model transmission near elimination because we can make a few assumptions that are not generally true. (1) There are many fewer infected than susceptible people and so the force of infection is independent of the susceptible fraction. (2) To know if elimination does not occur, we only need to predict the next paralytic case and do not need to correctly model outbreak size or duration. (3) Since we do not need to model the non-linear dynamics during an outbreak, but we want to explore the wide range possible transmission histories that could occur between cases or prior to elimination, a flexible statistical model of the force of infection is more appropriate than a specific dynamical model with mechanistic assumptions.

\begin{figure}[ht]
\includegraphics{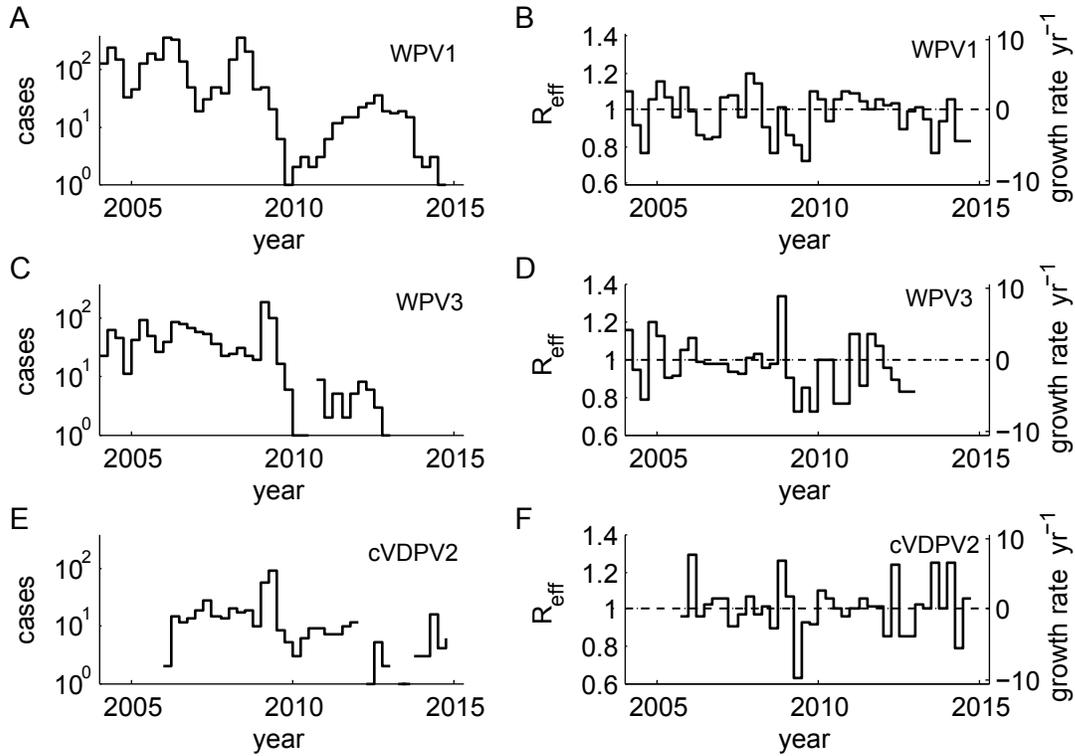}
\caption{{\bf Historical polio incidence in Nigeria.} Polio case counts in 3 month bins: WPV1 (A), WPV3 (C), cVDPV2 (E). Inferred growth rate (and reproductive number for $\mu = 26 \, \text{yr}^{-1}$): WPV1 (B), WPV3 (D), cVDPV2 (F). Note that while the overall case counts vary over time by two orders of magnitude, the reproductive number mean and range are stable. This motivates the assumption that transmission dynamics are stable among the populations that support ongoing transmission even as the number of such populations is lower now than a decade ago.}
\label{caseFig}
\end{figure}

\paragraph*{Statistical model for the time-varying force of infection.}
A statistical model of the force of infection was constructed from published polio incidence data (\nameref{S1}). 
The paralytic polio cases were re-binned to count the number of cases, $C(t)$, in $\Delta t = 3$ month intervals and the growth rate between intervals was estimated as $\lambda(t+\Delta t)-\mu=\frac{1}{\Delta t}\log\left(\frac{C(t+\Delta t)}{C(t)}\right) $. When there are no cases in an interval, the growth rate was assumed to be the same as in the nearest bin for which there was data (Fig. \ref{caseFig}).

\begin{figure}[ht]
\includegraphics{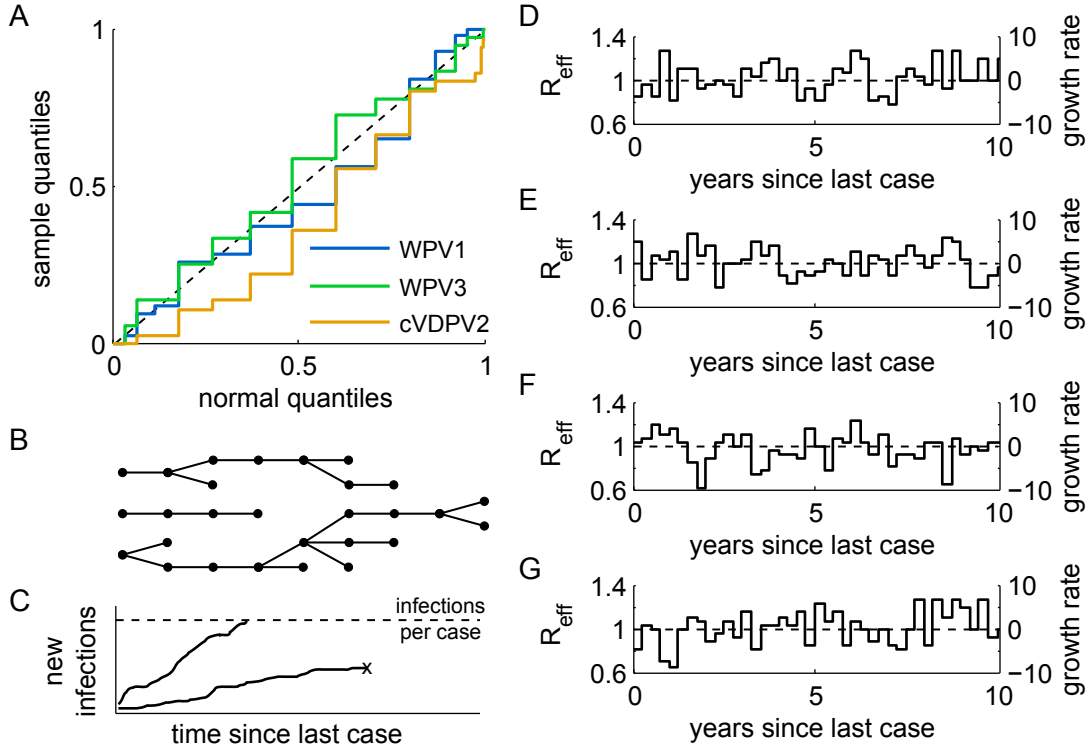}
\caption{{\bf Description of the model.} (A) Quantile-quantile plot of the growth rate distributions for each serotype (Fig. \ref{caseFig} B,D,F) vs a normal distribution with mean 0 and standard deviation $3.59 \, \text{yr}^{-1}$. There are no significant differences between serotypes or deviations from normality (KS-test, $p>0.05$). (B) Dynamical model cartoon: the model simulates the fade out and persistence of chains of transmission among a small number of people in a larger population. (C) Example model trajectories. Simulations were terminated when either enough infections were created to cause a new case or when elimination occurred. (D-G) Example simulated growth rate trajectories for the model (also indicated is $R_{\text{eff}}$ for $\mu=26\,\text{yr}^{-1}$). Compare with Fig. \ref{caseFig} B,D,F. }
\label{modelSpecFig}
\end{figure}

The distribution of growth rates is approximately normal. There are no significant differences by serotype (Fig. \ref{modelSpecFig}A) or significant deviations from normality (KS-test, $p>0.05$). The mean growth rate is $-0.38 \,\text{yr}^{-1}$ and insignificantly different from 0; the standard deviation is $3.44 \,\text{yr}^{-1}$\edit{; there is no significant dependence of the growth rate with time.} It is more intuitive to consider the effective reproductive number. For a mean infectious duration of 2 weeks, $\mu=26\,\text{yr}^{-1}$, the estimated mean effective reproductive number is 0.99 as expected for an endemic disease, and the standard deviation is 0.14. To test for temporal correlations indicative of seasonality, I examined the Fourier power spectra of each growth rate time series. The spectrum for WPV1 indicates the existence of seasonality with peak power at a 1 year period, but there are no clear peaks for WPV3 or cVDPV2 (data not shown). Despite decreases in polio prevalence and increases in vaccination quality since 2004 \cite{Upfill-Brown2014}, the observed growth rate mean and range appear to be stable.

\edit{The estimator is insensitive to changes in the surveillance quality over time as long as the fraction change in surveillance sensitivity, $\delta s$, in the estimation time interval is small. The growth rate estimator with a term for the unknown changing surveillance sensitivity is: $\frac{1}{\Delta t}\log\left(\frac{(1+\delta s) C(t+\Delta t)}{C(t)}\right) \approx \frac{1}{\Delta t}\log\left(\frac{C(t+\Delta t)}{C(t)}\right) + \frac{\delta s}{\Delta t}$.
The typical size of the first term is the standard deviation of the growth rate, and so as long as $\delta s \ll 0.86$, the change in surveillance sensitivity can be ignored. The national non-polio AFP detection rate has roughly doubled since 2005 \cite{WHO2007,Etsano2014}. Thus, while the slow increase in surveillance sensitivity is significant over the ten year period, its impact on the statistical model for the force of infection is negligible ($\delta s\approx 0.03$ per quarter).}

To model possible trajectories for the time-varying force of infection, I constructed time series for the growth rate $\lambda(t)-\mu$ from independent samples for each 3 month interval from the growth rate distribution. The assumption of independence between intervals is justified because weak seasonality has little influence on the interval between cases (Fig. 4 of ref. \cite{Kalkowska2012}). Example trajectories are shown in Fig. \ref{modelSpecFig}. The mean infectious duration was drawn from a uniform prior for $\mu$ on the interval $\mu=\left[(30 \,\text{days})^{-1},\,(7 \,\text{days})^{-1}\right]$. The range of the prior is determined from the range of observed mean poliovirus shedding durations in various historical and immunological settings \cite{Alexander1997}.  

This statistical model represents changes in the balance of vaccination coverage, demographic dynamics, and relevant heterogeneities as assumptions about the mean, variance, and temporal correlation of the force of infection. It is based on the assumption that there are no long-term trends in the force of infection. This assumption is conservative in the sense that trends in either direction shorten duration of silent transmission by either making outbreaks or elimination more likely. Finally, note that this model is designed to simulate disease dynamics near elimination. The model would not work well if the saturating nonlinearities that determine outbreak size and duration were important because the model is inherently linear. 

\paragraph*{Model of the interval between cases.}
To model the interval between cases, simulations were initialized at the time of the most recent observed case. I assumed that progression from infection to paralysis can be modeled as a Bernoulli trial with probability $p=(\text{case-to-infection})$.  Accordingly, the probability that infection number $N_i$ is the next paralytic case is geometrically distributed, $P(N_i;p)=p(1-p)^{N_i-1}$ \cite{DeGroot1986}. Estimated case-to-infection ratios\edit{ for individuals fully susceptible to paralysis} are 1:200 for WPV1, 1:1150 for WPV3, and 1:1900 for cVDPV2 \cite{Nathanson2010} under the assumption that cVDPV2 has an indistinguishable phenotype from wild type 2 \cite{Kew2005a}.\edit{ The impact of imperfect AFP surveillance with only $50\%$ sensitivity (such that half of all polio cases can be missed on average) was modeled by reducing the case-to-infection ratios to 1:400 for WPV1, 1:2300 for WPV3, and 1:3800 for cVDPV2 .} 

\edit{Since the case-to-infection ratios above describe paralysis-susceptible infections only, and the statistical model for the force of infection is based on paralytic cases only, the model of new infections explicitly represents only infections in the cohort of people who are susceptible to paralysis. It does not represent infections in individuals that are protected from paralysis by prior immunity but are not fully protected from infection \cite{Nathanson2010}. The model is agnostic to the details of how transmission between paralysis-susceptible individuals is mediated, and the role of individuals who can be infected but cannot be observed is subsumed into the empirical model for the force of infection between paralysis-susceptible individuals.}

\paragraph*{Initial condition distribution.}
The interval between the two most recent observed cases provides a target to calibrate the initial prevalence. The range of initial infections was calibrated to match the median interval between cases in the model to the observed interval between the two most recent cases. This is based on the assumption that prior to observing a silent period, the next case interval is expected to be similar to the previous one. 

For WPV1, the most recent case occurred on July 24, 2014 and the observed interval to the case prior to it was $\Delta t_c =58$ days earlier \cite{GPEI2015b}. (For comparison, since the start of 2012, the typical interval between WPV1 cases was $5.5 \,[1,27]$ days (mean [$95\%$ interval]).) For WPV3, the most recent case occurred on November 11, 2012 and the interval to the previous was $\Delta t_c=120$ days \cite{GPEI2015b} (vs. $15 \,[1,42]$ days). For cVDPV2, the most recent case occurred on November 16, 2014 and $\Delta t_c=13$ days \cite{GPEI2015b} (vs. $22 \,[1,141]$ days). 

For each $\mu$, gradient descent was used to reduce the absolute difference between the median time to the next case in the model and the target interval to at most one day. The resulting fits of the initial conditions to the median intervals are closely approximated by a simple curve that depends on $\mu$, the case-to-infection ratio for each serotype, the target interval between cases, and a factor of order 1: $$N_i(0)\approx f\frac{\left(\text{case-to-infection}\right)^{-1}}{\mu\Delta t_c}, $$ with $f_{\text{WPV1}}=0.55$, $f_{\text{WPV3}}=0.60$, and $f_{\text{cVDPV2}}=0.90$. 

The initial condition curves indicate that at the time of the most recent case, the prevalence of cVDPV2 was approximately 22 times greater than the prevalence of WPV1 and 73 times greater than WPV3 at the times of their most recent cases.

{\bf Added July 10, 2015: Bayesian data assimilation for updated cVDPV2 initial conditions.}
For the updated cVDPV2 forecast, a Bayesian data assimilation procedure was used to identify appropriate initial conditions. From the set of simulations from the March forecast that are consistent with the data as described in the Results, I built a mixture distribution for the initial number of infections on May 16, 2015:
\begin{equation*}
P\!\left(N_i(0)\big|\mu\right)=\displaystyle \sum_{\text{scenario}} P\!\left(\text{scenario}\right)P\!\left(N_i(0)\big|\text{scenario},\mu\right),
\end{equation*}
with 
\begin{equation*}
P\!\left(\text{scenario}\right) =
\begin{cases}
P\!\left(\text{blue}\right)=\frac{8}{35}, &\\
P\!\left(\text{green}\right)=\frac{23}{35}, &\\
P\!\left(\text{orange}\right)=\frac{4}{35}, &
\end{cases}
\end{equation*}
and normally-distributed initial conditions given the scenario
\begin{equation*}
P\!\left(N_i(0)\big|\text{scenario},\mu\right) =
\begin{cases}
P\!\left(N_i(0)\big|\text{blue},\mu\right)\sim\mathcal{N}\!\left(6\frac{1900}{181 \mu }, 200 \right), &\\
P\!\left(N_i(0)\big|\text{green},\mu\right)\sim\mathcal{N}\!\left(0.55\frac{1900}{181 \mu }, 65 \right), &\\
P\!\left(N_i(0)\big|\text{orange},\mu\right)\sim\mathcal{N}\!\left(0.09\frac{1900}{181 \mu }, 25 \right). &
\end{cases}
\end{equation*}

\paragraph*{Simulation procedure.}
The stochastic version of the model was coded in C\# and trajectories were simulated using the Gillespie's direct method for time-varying rates \cite{Gillespie1976,Anderson2007}. For each run of the model, the parameters $\mu$, $\lambda(t)$, and the number of infections required to produce the next case were sampled from their distributions, and the simulation was stopped when elimination occurred or the number of infections required to produce a case was reached. For all scenarios, at least 20~000 simulations and up to 200~000 simulations were run to produce Fig. \ref{resultsFig}. 

\paragraph*{Probabilities of elimination and the time to the next case.}
The probability of elimination without seeing another case was estimated as:
\begin{equation*}
P_{\text{elimination}}(t)=\frac{n_e(t)}{n_e(t) + n_p(t)},
\end{equation*}
where $n_e(t)$ is the number of simulations that eliminate at or before time $t$ and $n_p(t)$ is the number of simulations that persist past time $t$. The probability of the time to the next case given elimination does not occur and given the observed silent period through March 31, 2015 is estimated as the number of simulations that produce a case after March 31, 2015 but at or before time $t$ over the number of simulations that produce a case at any time after March 31, 2015. 
{\bf Added July 10, 2015:} For the updated cVDPV2 forecast, The probability of the time to the next case given elimination does not occur is estimated as the number of simulations that produce a case after May 16, 2015 but at or before time $t$ over the number of simulations that produce a case at any time after May 16, 2015. 

\section*{Supporting Information}
\subsection*{S1 Table}
\label{S1}
{\bf Polio case incidence data table for 2004--2014.} The polio case incidence rates for for WPV1 and WPV3 through September 2014 were scraped from the figures of refs. \cite{WHO2007,WHO2008,WHO2010,WHO2012,Etsano2013,Etsano2014} with the aide of the DataThief software package (B. Tummers, DataThief III. 2006:\url{ http://datathief.org/}). The cVDPV2 incidence data was taken from the supplement of ref. \cite{Burns2013a} through the end of 2011 and extended through September 2014 with scraped data from refs. \cite{Etsano2013,Etsano2014}. The remaining data after September 2014 was derived from the weekly tally sheets available at ref. \cite{GPEI2015b}. 

\section*{Acknowledgments}
The author would like to thank Kevin McCarthy, Hil Lyons, Guillaume Chabot-Couture, Hao Hu, and Philip Eckhoff (IDM) for helpful feedback at various points during this project, Dr. Faisal Shuaib (Federal Ministry of Health, Federal Republic of Nigeria)\sout{ comments on}\edit{ and the anonymous reviewers for comments that improved} the manuscript, and Christopher Lorton, Basil Bayati, Joshua Proctor, and Min Roh (IDM) for sharing their efficient and easy-to-use C\# compartmental model solver. 



\nolinenumbers

%
%
%


\end{document}